\documentstyle[12pt]{article}
\begin{document}
\bibliographystyle{[baron.wp]aip}
\baselineskip = 22.2pt plus 0.2pt minus 0.2pt
\lineskip = 22.2pt plus 0.2pt minus 0.2pt
\newcommand{\lton}{\stackrel{\large <}{\sim}}
\newcommand{\gton}{\stackrel{\large >}{\sim}}
\newcommand{\leqn}{\stackrel{\large <}{=}}
\newcommand{\geqn}{\stackrel{\large >}{=}}
\newcommand{\procent}{^o/_o}
\newcommand{\beq}{\begin{equation}}
\newcommand{\beqar}{\begin{eqnarray}}
\newcommand{\eeq}[1]{\label{#1} \end{equation}}
\newcommand{\eeqar}[1]{\label{#1} \end{eqnarray}}
\newcommand{\gfm}{{\rm GeV/Fm}^3}
\newcommand{\fsbmn}{\langle F_{\mu \nu} \rangle}
\newcommand{\fspmn}{\langle F^{\mu \nu} \rangle}
\newcommand{\asbm}{\langle A_\mu \rangle}
\newcommand{\aspm}{\langle A^\mu \rangle}
\newcommand{\epsi}{\vec{\epsilon}}
\thispagestyle{empty}
\pagestyle{myheadings}
\markboth
{\it  Kuo, Ray, Shamanna and Su / Isospin Lattice Gas Model and Nuclear matter
}
{\it  Kuo, Ray, Shamanna and Su / Isospin Lattice Gas Model and Nuclear matter
}

\begin{center}
{\bf Isospin Lattice Gas Model and Nuclear-Matter}

{\bf   Phase Diagram and Pressure-Volume Isotherms }
\footnote [1]{Work supported in part by US DOE Grant DE-FG02-88ER40388}
\\[2ex]

T.T.S. Kuo, S. Ray \footnote [2]{Alternate address: Institute for
Theoretical Physics, Stony Brook, NY}, J. Shamanna , R.K. Su \footnote [3]
{Permanent address: Department of Physics, Fudan University, Shanghai,
People's Republic of China } \\

Physics Department, State University of New York at Stony Brook,
Stony Brook, NY 11794-3800, USA
\\[2ex]

\today \\[3ex]
{\bf ABSTRACT}\\[2ex]
\end{center}
\begin{small}

We study a cubic lattice gas model for nuclear matter where each lattice
site can be either occupied, by one proton or one neutron, or unoccupied.
A nearest-neighbor interaction of the form
$ - \sum_{<ij>} J_{ij}\tau_{zi} \tau_{zj}$
is assumed.  Our model is an isospin-1 Ising model, with ${\tau_z}$
= (1,0,-1) representing respectively (proton, vacancy, neutron).
 A kinetic-energy term has been included in our model.
Under the Bragg-Williams mean field approximation our model exhibits
the existence of a dense phase (liquid-like) and a rare phase
(gas-like). The nuclear-matter p-v isotherms given by our model are discussed.

\end{small}
\vfill
\eject

It is  believed that nuclear matter should exhibit a quark
deconfinement phase transition at high temperatures (${k_B T\sim 200\; MeV}$)
and/or high densities ($ \rho \sim 10 \rho _0$), $\rho _0$ being the saturation
density.
At low temperatures (${k_B T\sim 15-20\; MeV}$) and low
densities (${\rho \sim \rho_0}$) a liquid-gas phase transition of
nuclear matter is expected
to take place \cite{na87}. The underlying theoretical frameworks treating
these two types of phase transitions have so far been very different.

For the former one generally uses lattice gauge QCD and Monte Carlo
simulation \cite{ko83}. The space-time structure here is a lattice.
But for the latter, it has been a long tradition to employ
standard many-body theories such as Hartree-Fock
approximation \cite{va84,jaq83}, real time Green's function
method \cite{su87,su89},
and ring-diagram summation method \cite {ya87}. The space-time structure
here is a continuous manifold. There appears to be a space-time disparity and
one would like to ask the following questions: Is it necessary for nuclear
systems to have
so very different space-time geometry structures, namely a lattice and a
continuum, in different energy and density domains? Can we set our theory on a
unified footing so that a common lattice space-time manifold
may be used to treat
not only phase transitions in high-temperature and high-density regions but
also those with low temperatures and low densities?

The above concern has motivated us to explore new schemes for studying
nuclear matter in the low-temperature and
-density regions. About forty years ago Lee and Yang \cite{le52} suggested
a model of lattice gas where gas atoms are seated on a lattice. They mapped
the problem of a lattice gas with one type of atom into an Ising model
for spin half particles. They succeeded in describing a liquid-gas
phase transition for  atomic systems.
 We would like to generalize this model to nuclear matter which consists
of two types of nucleons: proton and neutron. Here we consider a lattice
with each site either being vacant or occupied by a proton or a neutron.
One is tempted to associate Ising spin 1 to a proton, -1 to a neutron,
and 0 to a vacancy. A preliminary account of our approach has been
reported \cite{jaya94}.

  Nuclear forces may be understood based on
one boson exchanges . The intermediate bosons include isoscalar
bosons ($\omega,\sigma,...$) and isovector bosons ($\pi,\rho,...$).
In this way,  effective nucleon-nucleon(NN)
interactions may be written in terms of a number of
standard components such as the spin-spin term $\sigma \cdot
\sigma $ , isospin-isospin
term $\tau \cdot \tau $, and the spin-isospin
term $\sigma \cdot \sigma \; \tau \cdot \tau$.
If we average over the spin and spatial variables,
we would obtain an
effective interaction which depends on isospin  only. The radial part of
the NN interaction, to a reasonable approximation, can be represented by a
nearest-neighbor square well potential with a repulsive hard core.
Thus we are led to conjecture the following interaction hamiltonian for nuclear
matter:
\beq
H_{int}= - \sum_{<ij>} J_{ij}\tau_{zi} \tau_{zj}
\eeq{1}
where
\beqar
\tau_{zi}=\left\{ \begin{array}{rl}
 1 & {\rm for\; proton}  \\
 0 & {\rm for\;vacancy} \\
-1 & {\rm for\; neutron.}  \end{array} \right.
\eeqar{2}
Note the above  is a nearest-neighbor interaction, as indicated by the
summation index $<ij>$, namely interaction
exists only between adjacent neighbors.
 $J_{ij}$ are interaction strength parameters. In fact we allow
for two such parameters only; $J_s$  for pp and nn interaction pairs and
$J_d$ for pn pairs. Clearly the above is a spin 1 Ising model \cite{bl66,ca66}.

We rewrite eq.(1) as
\beq
H_{int}= -J_s(N_{++}+N_{--})+J_d N_{+-}
\eeq{3}
where $N_{++},N_{--}$... represent respectively the
nearest neighbor pairs of
proton-proton, neutron-neutron, etc. Note that in our model there is no
interaction between vacancy and nucleon.

Now we proceed to calculate the thermodynamic quantities of our
system.  To facilitate the calculation, we introduce the following
variables. Let N denote the total number of lattice sites,
and $N_+,N_- ,N_0$  denote respectively the number of proton, neutron
and vacancy sites. We introduce
\beq
R=\frac{N_0}{N},\;S=\frac{N_+ - N_-}{N},\; N=N_+ +N_- + N_0
\eeq{4}
where R and S represent the relative emptiness and proton-neutron asymmetry
respectively. Note that in our model the nuclear-matter density $\rho$
is proportional to (1-R).

The spin-1/2 lattice-gas models for atomic systems have been used with
remarkable success in describing phase transitions. But they seem to have not,
to our knowledge, been applied to calculate the pressure-volume isotherms.
It has been suggested \cite{hu63} that in order to describe these isotherms
 one needs to add the ideal-gas pressure to the lattice gas grand potential.
The ideal gas pressure comes from the kinetic energy. Hence it would seem to be
preferrable to include the the kinetic energy in the hamiltonian, from the
beginning, rather than adding the ideal-gas pressure to the grand potential.
The above consideration has motivated us to include a kinetic-energy term in
our model, namely we employ a hamiltonian for the lattice gas of the form
\beq
H_{gas}= -J_s(N_{++}+N_{--})+J_d N_{+-}+ N\kappa (1-R)^{5/3}
\eeq{7}
where, guided by  the Fermi gas model, we have assumed the kinetic energy
per particle to be proportional to $\rho ^{2/3}$.
$\kappa$ is a constant which we shall discuss later.
The introduction of the
kinetic-energy term is very important for our model, as we shall see soon.

As an initial investigation let us adopt the
Bragg-Williams mean field approximation\cite{hu63}, namely
\beq
\frac{N_{+}^2}{N^2}\simeq\frac{N_{++}}{N \gamma  /2},\;
\frac{N_{-}^2}{N^2}\simeq
\frac{N_{--}}{ N \gamma /2},\; \frac{N_{0}^2}{N^2}\simeq\frac{N_{00}}
 { N \gamma /2}
\eeq{8}
where  $\gamma$ denotes the number of nearest neighbors of any given site,
and $ N\gamma /2$ is the total number of pairs.
For three dimensional simple cubic lattice, $\gamma =6$. From
 the constraints
$\gamma N_+ = 2N_{++} +N_{+-} +N_{+0}$,
$\gamma N_- = 2N_{--} +N_{+-} +N_{-0}$,
$\gamma N_0 = 2N_{00} +N_{+0} +N_{-0}$
and
$N=N_+ +N_- +N_0$, we obtain
\beq
N_{+-}=\gamma (N_+ +N_-) -(N_{++}+N_{--}) +N_{00} -\gamma N/2.
\eeq{6}
 We can now rewrite our hamiltonian of eq.(5) as
\beq
H_{gas}(R,S,N) =  -C_{1}NS^{2}-C_{2}N(1-R)^2 + N\kappa (1-R)^{5/3}
\eeq{9}
where
\beq
C_{1} =\frac {(J_s+J_d) \gamma }{4},\;C_{2}=\frac {(J_s-J_d)\gamma }{4}.
\eeq{10}

The grand partition function of our system is
\beqar
Q_G = \sum_{R,S} g(R,S,N)exp(-\bar{H}_{gas}/k_{B}T),\nonumber \\
\bar{H}_{gas}=H_{gas}(R,S,N)-hNS-\lambda N(1-R)
\eeqar{11}
 The multiplicity factor \cite{will} $g$ is
\beq
g(R,S,N)=\frac{N!}{N_{0}!N_{+}!N_{-}!} = \frac{N!}
{(NR)![N(1-R+S)/2]![N(1-R-S)/ 2]!}.
\eeq{12}
In $\bar{H}_{gas}$ we have included two Lagrange multipliers, h and $\lambda$.
In our grand partition function, we sum over all possible R and S values. Thus
we  have neither a definite number of nucleons nor a definite proton-neutron
assymmetry. The role of these Lagrange multipliers is to control the average
values of R, denoted as $\bar R$, and  of S, denoted as $\bar S$.
For instance for symmetric nuclear matter we need  $\bar S$=0, and this may
be attained by varying h. Similarly, different $\bar R$ values may be
obtained by varying $\lambda$.

Since our final goal is to study the thermodynamic limit
($N\rightarrow\infty$), we can replace the sum in the partition function
$Q_G$ by its most dominant term \cite{will} (assuming the dominant term to be
non-degenerate). Using Stirling's formula one obtains

\beqar
-k_B T\ln Q_{G} &=&  -C_{1}NS^2 -C_{2}N(1-R)^2 -h N S  \nonumber \\
& & - \lambda N (1-R) + \kappa N(1-R)^{5/3} \nonumber \\
& & + Nk_BT[R\ln R +\frac{(1-R+S)}{2}\ln(1-R+S) \nonumber \\
& & + \frac{(1-R-S)}{2}\ln(1-R-S) -(1-R)\ln 2]
\eeqar{13}
with the subsidiary extremum conditions
\beqar
\frac{\partial (-k_B T\ln Q_{G})}{\partial S}&=&\frac{Nk_B T}{2}
\ln \frac{1-R+S}{1-R-S} - [2C_{1}S + h] N = 0, \\
\frac{\partial (-k_B T\ln Q_{G})}{\partial R}&=&\frac{Nk_B T}{2}
\ln \frac{4R^2}{(1-R)^{2}-S^{2}}+2NC_{2}(1-R)  \nonumber \\
& & \mbox{}    \;\;\;       -\; \frac{5}{3} \kappa N(1-R)^{2/3} + \lambda = 0.
\eeqar{14}
It is seen that  $h = 0$ and $S=0$ is a special solution which
represents a symmetric nuclear matter. In this case eq.(13) becomes an
identity and eq.(14) reduces to
\beq
k_B T \ln \frac{2R}{1-R}+2C_{2}(1-R)- \frac{5}{3} \kappa (1-R)^{2/3}+
\lambda = 0.
\eeq{15}

The analysis of this equation gives us the most important results of
our present paper. It shows that below a certain temperature $T_c$
we have two phases, one is dense (liquid-like) and the other is
rare (gas-like). Above this temperature there is only one phase.
The existence of a liquid-gas phase transition together with the
 determination of its phase diagram
on the basis of a simple model that assumes only a phenomenological
two body, nearest-neighbor interaction is quite remarkable. It seems to
call for serious attention and further study.

 Let us consider the $\lambda$=0 case first. In this case we rewrite eq.(15)
as
\beqar
\chi (R,T)=f(R)-g(R,T)=0,\hskip 5cm \nonumber \\
f(R) \equiv \ln \frac{2R}{1-R};\hskip 0.4cm
g(R,T) \equiv \frac{5 \kappa}{3 k_B T} (1-R)^{2/3} -
\frac{2 C_{2}}{k_B T} (1-R).
\eeqar{17}
We note that f(R) is a monotonically increasing,
unbounded (at $R\rightarrow0+\;,\;\;$ and $R\rightarrow1- \;$)
function of R having one point of inflection at $R = 1/2$.
However g(R,T) is a bounded function in the same domain.
Hence g(R,T) must intersect f(R) at least once, i.e., eq.(16) must have at
least one solution. In addition g(R,T) has a negative curvature and one maximum
for $R\in[0,1]$. Thus there is a possibility of having more than one point
of intersection with f(R) below a suitable temperature $T_c$.

In Fig. 1 we display some typical behaviours of f(R) and g(R,T). As shown,
f(R), denoted by the solid line, approaches to $-\infty$ at R=0 and to
$\infty$ at R=1. g(R,T) is a concave-downward curve.
f(R) intersects the R axis at $R=1/3$.
g(R,T) intersects the R axis at two points: One is at  R=1 which is independent
of the values of the parameters $C_2,\; \kappa,$ and $k_BT$, and the second
point of intersection depends on the ratio $\alpha \equiv 5 \kappa/6 C_2$ and
importantly is independent of temperature T.

For low temperatures, the curves f(R) and g(R,T) may have three
intersection points, as denoted by A,B, and C for the T=8 case. (Note that
we use the convention of $k_B \equiv 1$.)
It is readily checked that the middle intersection point, i.e. B,
corresponds to
a minimum of $\ln Q_G$ and hence it is not a physical solution.
The intersection points A and C are the physical solutions.
Note that g(R,T) is inversely proportional to the temperature. Hence as T
increases, the right side of it sweeps down while its left side sweeps up, as
shown in the figure. At some critical temperature
T$_c$ the curves f(R) and g(R,$T_c$) become tangent to each other. And
afterwards the curves f(R) and g(R,T) have only
one intersection point. g(R,T) eventually becomes a horizontal line as T
approaches $\infty$, as
indicated by the $T=\infty$ dotted line in the figure, and it obviously
has just one intersection point with f(R).

In Fig. 2 we plot the solutions of eq.(16). As seen, we have two solutions,
as indicated by points A and C of Fig. 1, for
$T < T_c$, and  for $T > T_c$ we have only one solution. This figure
suggests that below T$_c$ we have three regions, the dense(liquid-like)
phase, the rare(gas-like) phase and the coexistence phase in between.
The physical meaning of the above results may become clearer by examining the
 pressure-volume isotherms.

Before doing this, we should probably discuss
 the parameters $C_2$ amd $\kappa$ which have entered
into our calculations. To have an attractive nearest-neighbor interaction,
we have $J_s >0$ and $J_d <0$. In this case  $C_2$ is positive.
The magnitude of $J_s$ and $J_d$ is probably comparable to the average
potential energy in nuclear matter, which is about -40 MeV. Hence we have
picked $C_2$ = 125 MeV, recalling that $C_2$ has been defined in eq.(9).
The parameter $\kappa$ may be estimated from the average kinetic energy
given by the Fermi gas model. Assuming a lattice spacing of 1.5 fm
and taking the nucleon mass as 940 MeV, $\kappa$ is obtained as about
125 MeV. Hence as a general guideline, the parameters $C_2$ and $\kappa$
should both be not too far from 125.
The results of Figs.1 and 2, and 3 to be presented later,
are all obtained with $C_2$=125 and $\kappa$=$1.2\alpha C_2$
with a specially chosen $\alpha$ as discussed below.

Near the critical point, there is a subtle dependence of the
solutions of eq.(16)  on the ratio  $\alpha \equiv \frac{5\kappa}{6C_2}$.
As indicated by points
A, B and C of Fig. 1, we have three intersection points below $T_c$. It is
readily checked that by choosing $\alpha =(2/3)^{1/3}$, these three
intersection points all merge together at the critical point.
Then the phase diagram
near $T_c$ has the smooth shape as shown in Fig. 2. If one uses a slightly
different ratio,  the merging would generally take place in two steps, first
involving two intersection points and then the third. This will lead to a phase
diagram with a "cusp" shape near the critical point, which may seem to be
rather unconventional. By intuition we feel that it is more reasonable to have
a smooth phase boundary, and hence we have chosen the above $\alpha$. With
this $\alpha$ and $C_2$=125, we have $\kappa$=131.037, which has been
used in obtaining the results presented in our Figs. 1 to 3.

For fixed T, the solutions of eq.(16) determines the R values where
the grand partition function has a maximum contribution. In fact this maximum
term is an overwhelmingly dominant one. (For finite lattices such as that with
N=$10^6$, numerical simulation has shown that the magnitude of
the maximum term is typically $\sim 10^{100}$ while for all the other
terms  it is $\sim 10^{20}$.)
Hence the R values given by eq.(16) are just the average values of R,
as denoted previously by $\bar R$, for the system at temperature T
and $\lambda$=0. The $\bar R$ values for  $\lambda \neq 0$ are given by
the solution of eq.(15). Then the pressure at various densities and
temperatures is given by
\beqar
p(\bar R,T) \equiv \frac{k_BT}{N} \ln Q_{G} \hskip 7cm \nonumber \\
 = C_2(1-\bar R)^2 - \kappa (1-\bar R)^{5/3} - k_BT \bar R\ln
(\frac{2 \bar R}{1-\bar R})
 - k_BT \ln (\frac{1-\bar R}{2})+\lambda (1-\bar R).
\eeqar{18}
Let us define the specific volume v
as $(1-\bar R)^{-1}$. Then we can calculate the p-v isotherms using
eqs.(16) and (18). Our results are shown in Fig. 3.

Our figure seems to contain some interesting features.
For a given temperature $T$ the isotherm is
obtained as a parametric plot of specific volume versus pressure with
$\lambda$ its generating parameter.
For $ T \geq T_c $ one gets a single smooth curve by varying
$\lambda$. The boundary, drawn as a solid line at v$\simeq 1.5$, corresponds to
$\lambda = 0$. The isotherm to the left of this boundary is obtained with
positive $\lambda$, while to the right of this boundary the isotherm
is obtained with negative $\lambda$. The critical temperature obtained
is $T_c = 18.4$.

For T below $T_c$, it is of  interest that no
isotherms are obtained in the intermediate region, i.e. region I.
 For  $\lambda < 0 $, we get a gas-like isotherm
 starting from the high specific volume tail of the phase boundary.
  And for  $\lambda > 0$ we get a liquid-like isotherm (very
high compression modulus) starting from the lower specific volume edge of
the phase boundary. The above is because when $T<T_c$, eq.(15) has no
solutions in region I, the coexistence region.  With the introduction of an
infinitesimal  $\lambda$ the system chooses one of the two values of
$\bar R$ admissible for the given $T < T_c$, depending on the sign of
$\lambda$.
 This phenomena is reminisent of the spontaneous symmetry breaking in
ferromagnetism.

There is another point which may be mentioned. As indicated in the figure,
for a given isotherm the pressure at the liquid boundary and that at the
gas boundary appear to be equivalent to each other, such as the
apparent equivalence between the pressures
of the T=12 isotherm at the two boundaries. We have examined this type of
apparent equivalence, and have found that these two pressures are "exactly"
equal to each other, within the accuracy of our computer. The structure
of our calculated isotherms seems to strongly support that for $T < T_c$
our model gives a liquid phase, a gas phase and a coexistence phase with the
boundary indicated by the solid line. And for $T> T_c$ the distinction
between the liquid and gas phases disappears.

As a conclusion, let us state the following points. We have studied a
simple lattice gas model for nuclear
matter, where each lattice site can be either vacant or occupied,
by one proton or by one neutron.
A hamiltonian consisted of a nearest-neighbor interaction and a
kinetic-energy term is assumed. The partition
function is then calculated with the
Bragg-William approximation. Some rather encouraging results have been
obtained. A phase diagram consisting of liquid, gas and coexistance phases
is obtained from our model. And the p-v isotherms given by our model are
surprisingly similar to those given by the van der Waals theory, except for the
difference that for $T < T_c$ our isotherms do not have the metastable
states in the coexistence region. Hence with our model one does not need
to determine the phase boundary by way of a Maxwell construction. The
liquid-gas critical
temperature obtained by us is $T_c = 18.4$ for symmetric nuclear matter,
 which is fairly close to the results given by earlier calculations
\cite{su87,jaq83}.

Comparing with earlier lattice gas models \cite{hu63}, a new ingredient of the
present model is the introduction of the kinetic energy term.
We recall that to have phase transitions
we need the g(R,T) curve to have three intersection points with f(R)
for $T< T_c$. Whether this happens or not clearly depends on $\alpha ~(=
5\kappa / 6C_2)$. There is a
wide range of values of $\alpha$, for which this could happen. But if we don't
have the kinetic energy term, then $\kappa=0$ and $\alpha = 0$. And in this
case the above requirement can not be met, and our model would  have no phase
transitions. Hence the inclusion of the
kinetic energy term is important for our model. In fact in a previous
paper \cite{su90} we have studied
a lattice-gas model for nuclear matter, using the same lattice Hamiltonian
as Eq.(5) but without the inclusion of the kinetic
energy term; the results were clearly  unsuccessful.

We have adopted a major approximation, the Braggs-Williams approximation,
in the present work. The accuracy of this approximation remains to be
investigated. More accurate calculations may be performed, using for example
the Bethe-Peierls
approximation \cite{hu63}, or Monte Carlo simulations which have been
extensively in lattice-gauge and Ising-model calculations. With the
Bragg-Williams approximation, the calculated phase boundary near the
critical point can be either a smooth shape or a cusp shape. It should be of
interest to see what would be the phase-boundary shape  given by such
more advanced methods.

\vfill\eject

\vfill\eject
\begin{center}
{\bf{FIGURE\,\,\,CAPTIONS}}
\end{center}
\begin{description}
\item[Fig.1] Graphical solution of eq.(16).
\item[Fig.2] Nuclear matter phase diagram given by our model.
\item[Fig.3] Nuclear matter p-v isotherms.
\end{description}
\end{document}